\documentclass[conference]{IEEEtran}

\IEEEoverridecommandlockouts

\usepackage[linesnumbered,ruled,vlined]{algorithm2e}

\SetKwInput{KwInput}{Input}                
\SetKwInput{KwInit}{Initialization}                

\usepackage{cite}
\usepackage{amsmath,amssymb,amsfonts}
\usepackage{algorithmic}
\usepackage{graphicx}
\usepackage{pgf}
\usepackage{textcomp}
\usepackage{caption}
\usepackage{subcaption}
\usepackage{comment}
\usepackage{gensymb}
\usepackage{todonotes}
\usepackage{orcidlink}
\usepackage{tikz}
\usetikzlibrary{calc}
\usepackage{relsize}
\usepackage{xcolor}
\usetikzlibrary{positioning} 
\usepackage{pgf}
\usepackage{tikz}

\usepackage{siunitx}
\tikzset{fontscale/.style = {font=\relsize{#1}}
    }
\def\BibTeX{{\rm B\kern-.05em{\sc i\kern-.025em b}\kern-.08em
    T\kern-.1667em\lower.7ex\hbox{E}\kern-.125emX}}
\begin{document}

\title{Leveraging von Mises Message-Passing\\ for Massive MIMO Detection\\
\thanks{This work has been partially supported by the ANR under the “France 2030” program under grant NF-PERSEUS (ANR-22-PEFT-0004), and under grant WARM-M2M (ANR-24-CE25-2514).}
}

\author{
\IEEEauthorblockN{Sweta Suresh}
\IEEEauthorblockA{
\textit{Inria, INSA Lyon,}\\
\textit{CITI Lab -- UR3720}\\
Villeurbanne, France \\
\orcidlinkc{0000-0003-0458-8342}}
\and
\IEEEauthorblockN{Arthur Michon}
\IEEEauthorblockA{\textit{Université de Toulouse,} \\
\textit{CNRS, Toulouse INP, IRIT}\\
Toulouse, France \\
\orcidlinkc{0009-0007-8072-2191} }
\and
\IEEEauthorblockN{Charly Poulliat}
\IEEEauthorblockA{\textit{Université de Toulouse} \\
\textit{CNRS, Toulouse INP, IRIT}\\
 Toulouse, France \\
\orcidlinkc{0000-0001-6407-8841}}
\and
\IEEEauthorblockN{Maxime Guillaud}
\IEEEauthorblockA{
\textit{Inria, INSA Lyon,}\\
\textit{CITI Lab -- UR3720}\\
Villeurbanne, France \\
\orcidlinkc{0000-0002-4105-4537}}
\and
\IEEEauthorblockN{Claire Goursaud}
\IEEEauthorblockA{
\textit{Inria, INSA Lyon,}\\
\textit{CITI Lab -- UR3720}\\
Villeurbanne, France \\
\orcidlinkc{0000-0003-0971-9305}}
}

\maketitle

\begin{abstract}
Detection in massive multiple input multiple output (mMIMO) systems suffers from exponential complexity while using optimal decoders like maximum a posteriori (MAP).  We propose a belief propagation-based detector based on directional statistics, applicable to  mMIMO systems relying on PSK modulations. Thanks to a continuous relaxation of the PSK modulation to the unit circle, and to the use of von Mises parametric representations of the messages to obtain sparse representations of the (generally infinite dimensional) messages, the proposed method allows for approximate detection with a low complexity which does not depend on the PSK modulation order. Extensions of the algorithm to imperfect channel realizations are also present. We quantify the performance and complexity of the proposed approach and compare it with detection algorithms based on Gaussian approximation. 
\end{abstract}

\section{Introduction}

Massive multiple input multiple output (mMIMO) \cite{bjornson2017massive} is a key technology for modern wireless communications. By equipping mMIMO base stations (BS) with a large number of antennas, it can handle a large number of users that transmit on the same time/frequency resource; the increased spatial multiplexing brings massive gain in spectral and energy efficiency. However, this entails joint detection of transmitted symbols from a high-dimensional and sometimes ill-conditioned channel. 
This high dimensionality requires algorithms capable of achieving a satisfactory trade-off between performance and complexity.

The optimal maximum a posteriori (MAP) detection rapidly becomes untractable because it has exponential complexity in constellation size and the number of symbols transmitted.
Most of the algorithms introduced in the literature  overcome this issue by assuming that received symbols follow a Gaussian distribution; this relaxation leads e.g. to zero forcing (ZF) or linear minimum mean square error (L-MMSE) detection, which however is still computationally intensive for large systems since it requires the computation of the pseudo-inverse of the channel matrix.
Gaussian relaxation has also been used to define approximate iterative algorithms schemes such as Gaussian belief propagation (Ga-BP) \cite{bickson2009PhD}, expectation propagation (EP) \cite{minka2005divergence} or approximate message passing (AMP) \cite{montanari2012graphical}.
A good introduction to the application of these message passing algorithms to mMIMO detection and the relation between these three algorithms can be found in \cite{wu2014low}. 

In this paper, we propose a new iterative detection algorithm applicable to PSK-modulated mMIMO based on directional statistics \cite{mardia2009directional_statistics_book}, more specifically von Mises (vM) distributions.
The use of directional statistics for the physical layer was studied in \cite{analog_digital_bp} to decode a coded pulse amplitude modulation. 
Building on variational message passing  \cite{winn2005variational} and mean field BP \cite{riegler2012merging,jakubisin2016probabilistic}, a BP algorithm using von Mises distributions was introduced in \cite{Suresh_vMBP_TBM_IZS2026}  in order to decode tensor-based modulations for unsourced random access.

The paper is organized as follows. We first introduce the detection problem in mMIMO in section~\ref{sec:sec_model}. In section~\ref{sec_vM-BP_coherent}, we briefly review the properties of von Mises distributions, and derive a message-passing algorithm that leverage von Mises belief propagation (vM-BP), for coherent detection of PSK symbols for the case of perfect channel state information (CSI). Section~\ref{sec_imperfectCSI} addresses the case of imperfect CSI. In section~\ref{sec_performance}, we detail the performance and complexity of the proposed approach, and benchmark it against state-of-the-art methods. Finally, section~\ref{sec_conclusion} concludes the article.

\section{System Model and Problem Formulation}
\label{sec:sec_model}

We consider a MIMO system with $K$ single-antenna transmitters and a receiver equipped with $N$ antennas. At each channel access, every user transmits $J$ bits of information that are mapped to a discrete PSK constellation $\Omega = \{e^{j2\pi m/M},  m=0,\dots,M-1 \} $ of cardinality $M=2^J$ (M-PSK). For simplicity and without loss of generality, we assume that every user uses the same constellation. 
The symbol transmitted by user $k$ is denoted by $x^{(k)}, k = 1,\ldots,K$.
We assume a narrow-band fading channel, with gains denoted by $\mathbf{H} = [\mathbf{h}^{(1)},\dots ,\mathbf{h}^{(K)}]\in\mathbb{C}^{N\times K}$ where $\mathbf{h}^{(k)} \in \mathbb{C}^N$ is the channel gains between user $k$ and $N$ receive antennas. 
The received signal is hence given by
\begin{equation}
    \mathbf{y} = \sum_{k=1}^{K} x^{(k)}\mathbf{h}^{(k)} + \mathbf{w},
    \label{MIMO_rx_signal}
\end{equation}
where $\mathbf{y} \in \mathbb{C}^{N}$ and $\mathbf{w} \in \mathbb{C}^N$ is a complex circular additive white Gaussian noise (AWGN) with zero mean and covariance matrix $\sigma^2 \mathbf{I}_N$, where $\mathbf{I}_N$ denotes the identity matrix of size $N$.

Assuming that $\mathbf{H}$ is known, the optimal solution for mMIMO detection considering the MAP criterion is given by:
\begin{equation} \label{eq_MAP}
    \hat{\mathbf{x}} = \arg\max_{\mathbf{x} \in \Omega^{K}} p(\mathbf{x} | \mathbf{y}; \mathbf{H} ).
\end{equation}
We can write the corresponding vector factorization of a posteriori vector probability as 
\begin{align}
\label{eq:vector_factor}
 p(\mathbf{x} | \mathbf{y}; \mathbf{H}) & \propto p(\mathbf{y} | \mathbf{x};\mathbf{H}) p(\mathbf{x}) \nonumber\\
  & \propto p(\mathbf{y} | \mathbf{x};\mathbf{H}) \prod_{k=1}^{K} f_{X^{(k)}}(x^{(k)}),
\end{align}
where $p(\mathbf{y} | \mathbf{x}; \mathbf{H})$ is the likelihood associated with the observation model given by \eqref{MIMO_rx_signal} and $f_{X^{(k)}}(x)$ is the prior distribution for variable $X^{(k)}$ (usually assumed uniform over $\Omega$). Unfortunately, computing $\hat{\mathbf{x}}$ as per \eqref{eq_MAP} has prohibitive complexity in mMIMO systems, because of the cardinality of the search space given by $\left|\Omega^{K}\right| = 2^{JK}$, which scales exponentially with $K$ and $J$. To overcome this complexity, suboptimal criteria are classically considered based on linear block detection, such as L-MMSE. Alternatively, approximate message-passing based detection methods are often considered to iteratively compute approximations of the a posteriori marginals $p\left(x^{(k)} \mid \mathbf{y};\mathbf{H}\right)$, enabling approximate MAP symbol detection. In the MIMO context, capitalizing on the knowledge of the channel matrix $\mathbf{H}$, we can rewrite \eqref{eq:vector_factor} as
\begin{align}
 p(\mathbf{x} \mid \mathbf{y}; \mathbf{H}) & \propto \prod_{n=1}^{N} p\left(y_n\mid \mathbf{x}; \mathbf{H}\right) \prod_{k=1}^{K} f_{X^{(k)}}(x^{(k)}) \nonumber \\
 & \propto \prod_{n=1}^{N} f_n(\mathbf{x}; y_n,\mathbf{h}_n) \prod_{k=1}^{K} f_{X^{(k)}}(x^{(k)}),
 \label{eq:scalar_factor}
 \end{align}
where $f_n(\mathbf{x}; y_n,\mathbf{h}_n)\triangleq p\left(y_n \mid \mathbf{x}; \mathbf{H}\right) =  p\left(y_n \mid \mathbf{x}; \mathbf{h}_n\right)$, with $\mathbf{h}_n$ being the $n$-th row of $\mathbf{H}$. This factorization naturally lends itself to a bipartite factor graph representation \cite{KschischangTIT01}, for which latent variable nodes are associated with symbols $x^{(k)}$ and observed variable nodes are associated with antenna observations $y_n$. 
The receive antenna $n$ is associated with a factor node $f_{n}$ (also denoted function node), defined by the corresponding local likelihood. This factor node links the observed variable $y_n$ to the latent variables $x^{(k)}, k=1, \ldots, K$, that contribute to this observation through the channel.

Based on the factor graph representation, different message-passing strategies can be considered to further lower the computational complexity. A common approach is to constrain the exchanged messages, which consist in distributions describing the information about a variable aggregated from the neighboring nodes in the graph, to belong to the family of Gaussian distributions. Doing so, each message is entirely characterized by a small number of parameters, such as its mean and variance.
Therefore, message-passing no longer consists of transmitting complete distributions, but only their parameters.
Such parametric message-passing approaches on the factor graph associated to the factorization in \eqref{eq:scalar_factor} lead to scalar algorithms such as Ga-BP, EP, and AMP. 
Note that when we consider the vector factorization as in \eqref{eq:vector_factor}, applying message-passing rule updates on the corresponding factor graph leads to the vector counterpart of the previously cited algorithm, such as the vector EP (VEP) algorithm \cite{senst2011framework}. 

In this paper, we restrict our study to M-PSK modulations, for which all symbols lie on the unit circle. This motivates the use of directional instead of Gaussian distributions for the parametric messages, as explained next. 

\section{vM-BP detection for Coherent MIMO}
\label{sec_vM-BP_coherent}
\subsection{von Mises Distribution: A Brief Review}
\label{sec_intro_vM}
The von Mises distribution (also known as circular normal distribution) \cite{mardia2009directional_statistics_book} is a distribution defined on the unit circle $\mathcal{C} =  \{x \in \mathbb{C} \ \mathrm{s.t.} \ |x|=1 \}$ which belongs to the exponential family. 
Let $X \in \mathcal{C}$ be a random variable following a von Mises distribution with parameter $\eta \in \mathbb{C}$; by definition,  the associated probability density function (p.d.f.) is
\begin{equation}
\mathrm{vM}(x;\eta)=\frac{1}{2 \pi I_0(\lvert \eta \rvert)} \exp\big(\mathrm{Re}(\eta^*x)\big).
\end{equation}
$\eta$ can be written as $\eta=\kappa e^{j\mu}$, where 
 $\mu = \mathrm{arg}(\eta) \in [0,2\pi)$ and $\kappa = \lvert \eta \rvert \in [0,\infty)$ are respectively the mean direction and concentration parameters.
An interesting property of this family is that vM distributions are fully defined by their first moment \cite{mardia2009directional_statistics_book}
\begin{align}
    m_1(\eta)&\triangleq \mathbb{E}(X)=\int_{x\in\mathcal{C}}{x  \ \mathrm{vM}(x;\eta)} \mathrm{d}x \nonumber \\
     & =A(\kappa) e^{j\mu}= A(|\eta|)\frac{\eta}{|\eta|},
\label{eq:vm_moment}
\end{align}
where $A(\kappa)=\frac{I_1(\kappa)}{I_0(\kappa)}$, and $I_i(.)$ denotes the $i$-th modified Bessel function of the first kind. 

We now introduce a new iterative detection algorithm based on vM-BP \cite{Suresh_vMBP_TBM_IZS2026} to perform message-passing-based MIMO detection on PSK-modulated symbols. The PSK modulation $\Omega$ is a discrete subset of $\mathcal{C}$; vM-BP consists of relaxing the domain of the optimization problem \eqref{eq_MAP} from $\Omega^K$ to $\mathcal{C}^K$, to obtain simple parametric messages for the BP implementation.

\subsection{Factor Graph Representation}
\label{sec:factor_graph}
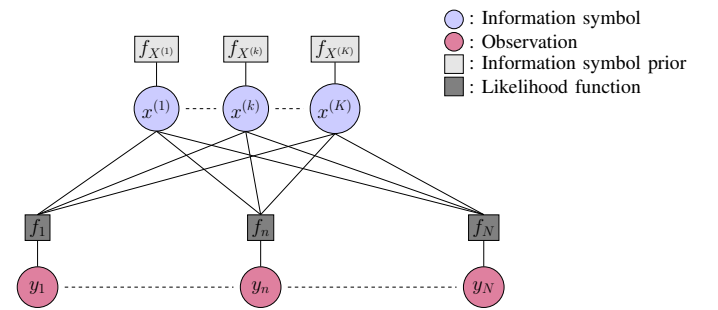
\begin{figure}[h]
    \centering
    \resizebox{0.5\textwidth}{!}{



\begin{tikzpicture}[scale=0.9]


\node[rectangle, draw, fill=black, fill opacity = 0.1, minimum size=16pt, text opacity = 1, fontscale=3] (PIN1) at (6,2) {$f_{X^{(1)}}$};
\node[rectangle, draw, fill=black, fill opacity = 0.1, minimum size=16pt, text opacity = 1, fontscale=3] (PIN2) at (9,2) {$f_{X^{(k)}}$};
\node[rectangle, draw, fill=black, fill opacity = 0.1, minimum size=16pt, text opacity = 1, fontscale=3] (PIN3) at (12,2) {$f_{X^{(K)}}$};


\node[circle, draw, fill=blue!20, minimum size=35pt, fontscale=3] (IN1) at (6,0) {$x^{(1)}$};
\node[circle, draw,fill=blue!20, minimum size=35pt, fontscale=3] (IN2) at (9,0) {$x^{(k)}$};
\node[circle, draw,fill=blue!20, minimum size=35pt, fontscale=3] (IN3) at (12,0) {$x^{(K)}$};

\draw[dashed, shorten <=2mm, shorten >=2mm] (IN2) to[bend left=0] (IN3);
\draw[dashed, shorten <=2mm, shorten >=2mm] (IN1) to[bend left=0] (IN2);

\draw (PIN1.south) -- (IN1.north);
\draw (PIN2.south) -- (IN2.north);
\draw (PIN3.south) -- (IN3.north);

\node[circle, draw, fill=purple!50,   minimum size=35pt, fontscale=3] (ON1) at (2,-6) {$y_1$};
\node[circle, draw, fill=purple!50,   minimum size=35pt, fontscale=3] (ON3) at (9.5,-6) {$y_n$};
\node[circle, draw, fill=purple!50,   minimum size=35pt, fontscale=3] (ON5) at (17,-6) {$y_{N}$};

\node[rectangle, draw, fill=black,   fill opacity = 0.5, minimum size=16pt, text opacity = 1, fontscale=3] (FON1) at (2,-4) {$f_1$};
\node[rectangle, draw, fill=black,  fill opacity = 0.5, minimum size=16pt, text opacity = 1, fontscale=3] (FON3) at (9.5,-4) {$f_n$};
\node[rectangle, draw, fill=black,   fill opacity = 0.5, minimum size=16pt, text opacity = 1, fontscale=3] (FON5) at (17,-4) {$f_{N}$};

\draw (FON1.south) -- (ON1.north);
\draw (FON3.south) -- (ON3.north);
\draw (FON5.south) -- (ON5.north);

\draw[dashed, shorten <=2mm, shorten >=2mm] (ON1) to[bend left=0] (ON3);
\draw[dashed, shorten <=2mm, shorten >=2mm] (ON3) to[bend left=0] (ON5);

\draw (FON1.north) -- (IN1.south);
\draw (FON1.north) -- (IN2.south);
\draw (FON1.north) -- (IN3.south);
\draw (FON3.north) -- (IN1.south);
\draw (FON3.north) -- (IN2.south);
\draw (FON3.north) -- (IN3.south);
\draw (FON5.north) -- (IN1.south);
\draw (FON5.north) -- (IN2.south);
\draw (FON5.north) -- (IN3.south);

\begin{scope}[shift={(16,3)}] 
    \node[circle, draw, fill=blue!20, minimum size=16pt] (leg1) {};
    \node[right=0.5mm of leg1,fontscale=3] {: Information symbol};


    \node[circle, draw, fill=purple!50, minimum size=16pt, below=1mm of leg1] (leg2) {};
    \node[right=0.5mm of leg2,fontscale=3] {: Observation};

    \node[rectangle, draw, fill=black,,   fill opacity = 0.1, minimum size=16pt, text opacity = 1 , below=1mm of leg2] (leg3) {};
    \node[right=0.5mm of leg3,fontscale=3] {: Information symbol prior};


    \node[rectangle, draw, fill=black,,   fill opacity = 0.5, minimum size=16pt, text opacity = 1 , below=1mm of leg3] (leg4) {};
    \node[right=0.5mm of leg4,fontscale=3] {: Likelihood function};

\end{scope}

\end{tikzpicture}

    }
    \caption{Factor graph representation for MIMO.}
    \label{factor_graph_MIMO}
\end{figure}

Let us first consider the coherent case where the channel is perfectly known. 
The system model described in Sec.~\ref{sec:sec_model} with the factorization of \eqref{eq:scalar_factor} can be represented by the factor graph given in Fig. \ref{factor_graph_MIMO}.
The graph contains two types of nodes: a) \textit{variable nodes} (VNs),  denoted as circular nodes, comprising information symbol nodes $x^{(k)}$ and observation symbol nodes $y_n$;  b) \textit{function nodes} (FNs), denoted as rectangular boxes. 
There are two types of function nodes, namely prior functions (denoted by $f_{X^{(k)}}$) and local likelihood functions attached to observations (denoted by $f_n$).
\paragraph{Priors} Starting with the prior functions, we assume that the true prior of information symbols is uniform on $\Omega$, given by $f_{X^{(k)}}(x) = \sum_{\alpha\in \Omega}\frac{1}{M} \delta(x-\alpha)$,
where $\delta$ is the Dirac delta function. 
Since vM-BP relies on the relaxation of the problem to the unit circle, we use instead a uniform prior on $\mathcal{C}$, which can be expressed as a von Mises distribution with concentration parameter equal to zero, i.e. $f_{X^{(k)}}(x) = \mathrm{vM}\left(x; 0\right)$. 

\paragraph{Observations}
The other type of function nodes consists of the local likelihood functions of the observations ${y_n}$ given $\mathbf{x}$ and $\mathbf{h}_n$, i.e.  
\begin{equation}
    f_n\biggl(\mathbf{x}; y_n ,\mathbf{h}_n\biggr)
        = \frac{1}{\pi \sigma^2}
        \exp{ \biggl\{ \frac{-1}{\sigma^2} \left |  y_n- \sum_{k}  x^{(k)}  h^{(k)}_n   \right| ^2 \biggr\} },
    \label{eq:likelihood_MIMO}
\end{equation}
where $n \in \{1,\dots,N\}$, $\mathbf{x} = [x^{(1)},\dots,x^{(K)}]$ and $\mathbf{h}_n = [h^{(1)}_n,\dots,h^{(K)}_n]$. Given this representation, there are $K$ symbols to be estimated based on $N$ observations.
Each information symbol node $x^{(k)}$ is connected to $N$ function nodes, while each function node $f_n$ is connected to $K$ information symbols. Thus, the total number of edges in the central part of the factor graph is equal to $KN$.

\subsection{vM-BP Messages and Beliefs}
In this subsection we define an approximate belief propagation algorithm on the factor graph represented on Fig. \ref{factor_graph_MIMO}, whereby
all messages to and from an information symbol node are constrained to belong to the parametric family of von Mises distributions.
This allows to replace a discrete probability distribution over $\Omega$ by a continuous distribution over $\mathcal{C}$ parameterized by a single complex parameter, as detailed in Section~\ref{sec_intro_vM}.
We show in the sequel that this yields parametric expressions of the messages which also belong to the family of von Mises distributions, and furthermore admit simple parameter update expressions. 
Let us now consider the two types of messages that occur in the graph at iteration $t$, and the resulting belief expressions.

\subsubsection{Message from VN $x^{(k)}$ to FN $f_n$} \label{message_from_x_to_f_sec}
It is given by
\begin{equation}
    \lambda^{[t]}_{{x^{(k)}  \rightarrow  f_{n}} }(x^{(k)})= \prod_{{n'\neq n}} \lambda_{f_{{n'}} \rightarrow x^{(k)}}^{[t-1]}(x^{(k)}),
    \label{message_from_variable_node_to_function_node}
\end{equation}
where $\lambda_{f_{{n'}} \rightarrow x^{(k)}}^{[t-1]}(x) \propto \textrm{vM}(\, x \,;\eta^{[t-1]}_{f_{n'}  \rightarrow x^{(k)} })$.
Therefore, the message $\lambda^{[t]}_{x^{(k)}  \rightarrow  f_{n}}$ is a product of exponential distributions which can be identified as a von Mises distribution $\lambda^{[t]}_{x^{(k)}  \rightarrow  f_{n}}(x) \propto \textrm{vM}(\,x\,;\eta^{[t]}_{x^{(k)}  \rightarrow  f_{n}})$, with natural parameter given as the sum of the natural parameters of the incoming messages at the previous iteration:
\begin{equation}
    \eta^{[t]}_{x^{(k)}  \rightarrow  f_{n}} =  \sum_{{n'\neq n}} \eta_{f_{{n'}} \rightarrow x^{(k)}}^{[t-1]}.
    \label{eq:natural_parameter_vn_to_fn}
\end{equation}

\subsubsection{Message from FN $f_n$ to VN $x^{(k)}$} \label{message_f_to_x_sec}
This is given by
\begin{equation}
\begin{split}    
   \lambda_{f_{n} \rightarrow x^{(k)}}^{[t]}(x^{(k)}) = 
   \int_{\mathbf{x}_{\sim x^{(k)}} \in \mathcal{C}^{K-1}} 
   & f_n\biggl( \mathbf{x} ; y_n,\mathbf{h}_n\biggr) \\
   & \prod_{k'\neq k} \lambda_{x^{({k'})} \rightarrow f_{n}}^{[t-1]}(x^{(k')}) d\mathbf{x}_{\sim x^{(k)}}, \\
\end{split}   
\label{eq:message_from_function_node_to_variable_node}
\end{equation}
where the integration is over the $K-1$ information symbols from all users but the $k^{\text{th}}$, denoted by the subscript $\sim x^{(k)}$.
Considering the definitions of the function node in \eqref{eq:likelihood_MIMO} and of the messages $\lambda^{[t]}_{x^{(k)}  \rightarrow  f_{n}}$, the integrand in \eqref{eq:message_from_function_node_to_variable_node} is a product of Gaussian and von Mises distributions, making the integral non-tractable. To overcome this issue, we make use of mean field approximation (see \cite{riegler2012merging} and \cite{jakubisin2016probabilistic}) and replace \eqref{eq:message_from_function_node_to_variable_node} by 
\begin{equation}
    \lambda_{f_{n} \rightarrow x^{(k)}}^{[t]} (x^{(k)}) =  \exp{\left(\mathbb{E}_{\sim x^{(k)}}[\log(f_n(\mathbf{x}; y_n ,\mathbf{h}_n))]\right)}.
\end{equation}
Taking the logarithm of the message yields
\begin{equation}
\begin{split}
   & \log \lambda_{f_{n} \rightarrow x^{(k)}}^{[t]}(x^{(k)}) = \\ & \frac{-1}{\sigma^2} \left| y_n- h_n^{(k)}x^{(k)} -\sum_{{k'}\neq k}m_1(\eta_{ x^{({k'})}\rightarrow f_n}^{[t-1]} )h^{({k'})}_n  \right| ^2,\\
   \end{split}
\end{equation}
where $m_1(\eta_{ x^{({k'})}\rightarrow f_n}^{[t-1]} )$ is the first moment of the message from interfering variable node $k'$ to factor node $f_n$, as per \eqref{eq:vm_moment}. After expanding and removing the constant terms with respect to $x^{(k)}$ by using the fact  $\rvert x^{(k)}\rvert^2=1$, the message to $x^{(k)}$ from function node $f_n$ is given by
\begin{equation}
\begin{split}
    & \lambda_{f_{n} \rightarrow x^{(k)}}^{[t]} (x^{(k)})\propto \\
     & \exp{\biggl\{ \mathrm{Re} \left( \frac{2h_n^{(k)}}{\sigma^2}\left(y_n^*  - \sum_{{k'}\neq k}m_1^*(\eta_{ x^{({k'})}\rightarrow f_n}^{[t-1]}) h^{*({k'})}_n  \right) x^{(k)} \right) \biggr\}}.\\
\end{split}
\label{eq:belief_fn_to_vn}
\end{equation}
This expression can be identified as a von Mises distribution with natural parameter 
\begin{equation}
   \eta_{f_{n} \rightarrow x^{(k)}}^{[t]} =  \frac{2(h_n^{(k)})^*}{\sigma^2}\left(y_n  - \sum_{{k'}\neq k}m_1(\eta_{ x^{({k'})}\rightarrow f_n}^{[t-1]}) h^{({k'})}_n  \right)
    \label{eq:natural_parameter_function_variable_final}.
\end{equation}

\subsubsection{Beliefs}
The beliefs of information symbol $x^{(k)}$ at iteration $t$  can be computed according to 
\begin{equation}
    b^{[t]}(x^{(k)}) \propto f_{X^{(k)}}(x^{(k)})  \prod_{n\in \{1,\dots,N\}} \lambda_{f_{{n}} \rightarrow x^{(k)}}^{[t]}(x^{(k)}).
    \label{belief_MIMO}
\end{equation}
Since every term in the product is a von Mises distribution, $b^{[t]}(x^{(k)}) $ is also von Mises, with natural parameter
\begin{equation}
    \hat{\eta}_{x^{(k)}}^{[t]} = \eta^{[0]}_{x^{(k)}} + \sum_{{n}} \eta_{f_{{n}} \rightarrow x^{(k)}}^{[t]},
    \label{eta_MIMO}
\end{equation}
where $\eta^{[0]}_{x^{(k)}}=0$ for the uniform prior.

\subsection{vM-BP Algorithm Outline}

At iteration 0 (initialization), there is no information but the prior, hence, 
$ \lambda^{[0]}_{x^{(k)}  \rightarrow  f_{n}}(x) = \text{vM}\left(\,x\,;0\right)$. Given this initial message, the first message from the function node ($f_n$) to the variable node ($x^{(k)}$) is 
$\lambda_{f_{{n}} \rightarrow x^{(k)}}^{[0]}(x^{(k)}) = \text{vM}\left(\,x^{(k)}\,;2(h_n^{(k)})^*y_n/\sigma^2\right)$.
The receiver then iteratively updates the messages in both directions using \eqref{eq:natural_parameter_vn_to_fn} and \eqref{eq:natural_parameter_function_variable_final} for a fixed number of iterations $N_{\mathrm{itr}}$, or until convergence. 
At the end of the algorithm, the information symbols are estimated from the beliefs \eqref{eta_MIMO}.
The MAP hard decision rule for the estimated PSK symbol is given by
\begin{equation} 
    \hat{x}^{(k)} = \arg\max_{x\in \Omega}  \mathrm{vM}(x;\hat{\eta}_{x^{(k)}}^{[N_{\mathrm{itr}}]}), \ \forall k=1,\cdots,K.
    \label{estimation_x}
\end{equation}

\section{vM-BP detection for MIMO with imperfect channel knowledge}
\label{sec_imperfectCSI}
The discussion so far focused on estimation of information symbols $x^{(k)}$ with perfectly known channel. Let us now consider a case where the receiver has an imperfect knowledge of the channel and tries to infer both the information symbols and the true channel state. Let $\tilde{h}_n^{(k)}$ denote the imperfect channel knowledge available to the receiver.
The prior of the channel variables captures the imperfect CSI through a complex Gaussian distribution given by 
\begin{equation}
\begin{split}
    f_{H_{n}^{(k)}}(h) & = \frac{1}{\pi \sigma_e^2} \exp{\biggl\{\frac{ -\lvert h -\mu_{h_n^{(k)}}^{[0]} \rvert ^2}{\sigma_e^2}\biggr\}}\\
    & = \text{CN}\left(h;\mu_{h_n^{(k)}}^{[0]},\sigma_e^2\right),
\end{split}
    \label{prior_pdf_h}
\end{equation}
with mean $\mu_{h_n^{(k)}}^{[0]} = \tilde{h}_n^{(k)}$ and variance $\sigma_e^2$. 

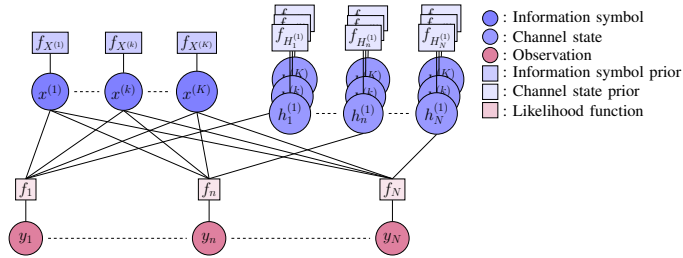
\begin{figure}[h]
    \centering
    \resizebox{0.5\textwidth}{!}{



\begin{tikzpicture}[scale=0.9]


\node[rectangle, draw, fill=blue!20, minimum size=16pt, text opacity = 1, fontscale=3] (PIN1) at (3,2) {$f_{X^{(1)}}$};
\node[rectangle, draw, fill=blue!20,  minimum size=16pt, text opacity = 1, fontscale=3] (PIN2) at (6,2) {$f_{X^{(k)}}$};
\node[rectangle, draw, fill=blue!20,minimum size=16pt, text opacity = 1, fontscale=3] (PIN3) at (9,2) {$f_{X^{(K)}}$};

\node[rectangle, draw, fill=blue!10, minimum size=16pt,fontscale=3] (PCN1) at (13,3,0) {$f_{H_1^{(K)}}$};
\node[rectangle, draw, fill=blue!10, minimum size=16pt,fontscale=3] (PCN2) at (13,2.7,0.25) {$f_{H_1^{(k)}}$};
\node[rectangle, draw, fill=blue!10, minimum size=16pt,fontscale=3] (PCN3) at (13,2.4,0.5) {$f_{H_1^{(1)}}$};
\node[rectangle, draw, fill=blue!10, minimum size=16pt,fontscale=3] (PCN4) at (16,3,0) {$f_{H_n^{(K)}}$};
\node[rectangle, draw, fill=blue!10, minimum size=16pt,fontscale=3] (PCN5) at (16,2.7,0.25) {$f_{H_n^{(k)}}$};
\node[rectangle, draw, fill=blue!10, minimum size=16pt,fontscale=3] (PCN6) at (16,2.4,0.5) {$f_{H_n^{(1)}}$};
\node[rectangle, draw, fill=blue!10, minimum size=16pt,fontscale=3] (PCN7) at (19,3,0) {$f_{H_N^{(K)}}$};
\node[rectangle, draw, fill=blue!10, minimum size=16pt,fontscale=3] (PCN8) at (19,2.7,0.25) {$f_{H_N^{(k)}}$};
\node[rectangle, draw, fill=blue!10, minimum size=16pt,fontscale=3] (PCN9) at (19,2.4,0.5) {$f_{H_N^{(1)}}$};

\node[circle, draw, fill=blue!50, minimum size=35pt, fontscale=3] (IN1) at (3,0) {$x^{(1)}$};
\node[circle, draw,fill=blue!50, minimum size=35pt, fontscale=3] (IN2) at (6,0) {$x^{(k)}$};
\node[circle, draw,fill=blue!50, minimum size=35pt, fontscale=3] (IN3) at (9,0) {$x^{(K)}$};

\node[circle, draw,fill=blue!40, minimum size=35pt,fontscale=3] (CN1) at (13.0000,0.5,0) {$h_{1}^{(K)}$};
\node[circle, draw,fill=blue!40,minimum size=35pt,fontscale=3] (CN2) at (13.0000,-0.1,0.25) {$h_{1}^{(k)}$};
\node[circle, draw,fill=blue!40,minimum size=35pt,fontscale=3] (CN3) at (13.0000,-0.7,0.5) {$h_{1}^{(1)}$};

\node[circle, draw,fill=blue!40, minimum size=35pt,fontscale=3] (CN4) at (16,0.5,0) {$h_{n}^{(K)}$};
\node[circle, draw,fill=blue!40,minimum size=35pt,fontscale=3] (CN5) at (16.0000,-0.1,0.25) {$h_{n}^{(k)}$};
\node[circle, draw,fill=blue!40,minimum size=35pt,fontscale=3] (CN6) at (16.00000,-0.7,0.5) {$h_{n}^{(1)}$};

\node[circle, draw,fill=blue!40, minimum size=35pt,fontscale=3] (CN7) at (19.000,0.5,0) {$h_{N}^{(K)}$};
\node[circle, draw,fill=blue!40,minimum size=35pt,fontscale=3] (CN8) at (19.0000,-0.1,0.25) {$h_{N}^{(k)}$};
\node[circle, draw,fill=blue!40,minimum size=35pt,fontscale=3] (CN9) at (19.0000,-0.7,0.5) {$h_{N}^{(1)}$};

\draw[dashed, shorten <=2mm, shorten >=2mm] (IN2) to[bend left=0] (IN3);
\draw[dashed, shorten <=2mm, shorten >=2mm] (IN1) to[bend left=0] (IN2);
\draw[dashed, shorten <=2mm, shorten >=2mm] (CN6) to[bend left=0] (CN9);
\draw[dashed, shorten <=2mm, shorten >=2mm] (CN3) to[bend left=0] (CN6);

\draw (PIN1.south) -- (IN1.north);
\draw (PIN2.south) -- (IN2.north);
\draw (PIN3.south) -- (IN3.north);
\draw (PCN3.south)+(0.2,0) -- (CN1.north);
\draw (PCN3.south)+(0.1,0) -- (CN2.north);
\draw (PCN3.south) -- (CN3.north);
\draw (PCN6.south)+(0.2,0) -- (CN4.north);
\draw (PCN6.south)+(0.1,0) -- (CN5.north);
\draw (PCN6.south) -- (CN6.north);
\draw (PCN9.south)+(0.2,0) -- (CN7.north);
\draw (PCN9.south)+(0.1,0) -- (CN8.north);
\draw (PCN9.south) -- (CN9.north);

\node[circle, draw, fill=purple!50,   minimum size=35pt, fontscale=3] (ON1) at (2,-6) {$y_1$};
\node[circle, draw, fill=purple!50,   minimum size=35pt, fontscale=3] (ON3) at (9.5,-6) {$y_n$};
\node[circle, draw, fill=purple!50,   minimum size=35pt, fontscale=3] (ON5) at (17,-6) {$y_{N}$};

\node[rectangle, draw, fill=purple!20,   fill opacity = 0.5, minimum size=16pt, text opacity = 1, fontscale=3] (FON1) at (2,-4) {$f_1$};
\node[rectangle, draw, fill=purple!20,  fill opacity = 0.5, minimum size=16pt, text opacity = 1, fontscale=3] (FON3) at (9.5,-4) {$f_n$};
\node[rectangle, draw, fill=purple!20,   fill opacity = 0.5, minimum size=16pt, text opacity = 1, fontscale=3] (FON5) at (17,-4) {$f_{N}$};

\draw (FON1.south) -- (ON1.north);
\draw (FON3.south) -- (ON3.north);
\draw (FON5.south) -- (ON5.north);

\draw[dashed, shorten <=2mm, shorten >=2mm] (ON1) to[bend left=0] (ON3);
\draw[dashed, shorten <=2mm, shorten >=2mm] (ON3) to[bend left=0] (ON5);

\draw (FON1.north) -- (IN1.south);
\draw (FON1.north) -- (IN2.south);
\draw (FON1.north) -- (IN3.south);
\draw (FON3.north) -- (IN1.south);
\draw (FON3.north) -- (IN2.south);
\draw (FON3.north) -- (IN3.south);
\draw (FON5.north) -- (IN1.south);
\draw (FON5.north) -- (IN2.south);
\draw (FON5.north) -- (IN3.south);
\draw (FON1.north) -- (CN3.west);
\draw (FON3.north) -- (CN6.south);
\draw (FON5.north) -- (CN9.south);

\begin{scope}[shift={(21,3)}] 
    \node[circle, draw, fill=blue!50, minimum size=16pt] (leg1) {};
    \node[right=0.5mm of leg1, fontscale=3] {: Information symbol};

    \node[circle, draw, fill=blue!40, minimum size=16pt, below=1mm of leg1] (leg2) {};
    \node[right=0.5mm of leg2, fontscale=3] {: Channel state};

    \node[circle, draw, fill=purple!50, minimum size=16pt, below=1mm of leg2] (leg3) {};
    \node[right=0.5mm of leg3, fontscale=3] {: Observation};

    \node[rectangle, draw, fill=blue!20, minimum size=16pt,  below=1mm of leg3] (leg4) {};
    \node[right=0.5mm of leg4, fontscale=3] {: Information symbol prior};

    \node[rectangle, draw, fill=blue!10,minimum size=16pt,  below=1mm of leg4] (leg5) {};
    \node[right=0.5mm of leg5, fontscale=3] {: Channel state prior};

    \node[rectangle, draw, fill=purple!20,minimum size=16pt, below=1mm of leg5] (leg6) {};
    \node[right=0.5mm of leg6, fontscale=3] {: Likelihood function};

\end{scope}







\end{tikzpicture}

    }
    \caption{Factor graph representation for MIMO with imperfect channel knowledge.}
    \label{factor_graph_MIMO_imperfect_chan}
\end{figure}

Fig.~\ref{factor_graph_MIMO_imperfect_chan} depicts the updated factor graph with an additional VN type for the channel state variables, denoted by $h_n^{(k)}$, and an additional FN type, denoted by $f_{H_n^{(k)}}$. In this representation, each channel state node is connected to a single FN, and observation FNs $f_n$ are connected to $2K$ VNs. In this factor graph, in Fig. \ref{factor_graph_MIMO_imperfect_chan}, there are four types of messages that can occur: messages to and from $x^{(k)}$ nodes; and messages to and from $h_n^{(k)}$ nodes. The messages to and from a channel node are constrained to be complex normal (CN) distributions; and the messages to and from an information symbol node are constrained to be vM distributions. The message from VN $x^{(k)}$ to FN $f_n$ remains the same as in Sec. \ref{message_from_x_to_f_sec}. The rest of the messages are discussed below. 

\subsection{Message from VN $h_n^{(k)}$ to FN $f_n$}
This is given by 
\begin{equation}
    \lambda^{[t]}_{h_n^{(k)}  \rightarrow  f_{n}}(h_n^{(k)}) = \prod_{n'\neq n} \lambda_{f_{{n'}} \rightarrow h_n^{(k)}}^{[t-1]}(h_n^{(k)}),
    \label{message_h_to_fn_imperfec_def}
\end{equation}
where $\lambda_{f_{{n'}} \rightarrow h_n^{(k)}}^{[t-1]} (h_n^{(k)}) \propto \text{CN}\left(h_n^{(k)}; \mu^{[t-1]}_{f_{n'}  \rightarrow h_n^{(k)} },\sigma^{2[t-1]}_{f_{n'}  \rightarrow h_n^{(k)} } \right)$ and $n'$ takes all values in $\{1,\cdots,N\}$ that are connected to the $h_n^{(k)}$ node except $n$. Since every term in \eqref{message_h_to_fn_imperfec_def} is a CN distribution, the resultant is also a CN distribution given as $\lambda^{[t]}_{h_n^{(k)}  \rightarrow  f_{n}}(h_n^{(k)}) = \text{CN}\left(h_n^{(k)}; \mu^{[t]}_{h_n^{(k)}  \rightarrow  f_{n}},\sigma^{2[t]}_{h_n^{(k)}  \rightarrow  f_{n}} \right)$, with parameters given as 
\begin{equation}
    \begin{split}
        & \sigma^{2[t]}_{h_n^{(k)}  \rightarrow  f_{n}} = \left(\sum_{n'}\frac{1}{\sigma^{2[t-1]}_{f_{n'} \rightarrow h_n^{(k)}}}\right)^{-1},\\
        & \mu^{[t]}_{h_n^{(k)}  \rightarrow  f_{n}} = \sigma^{2[t]}_{h_n^{(k)}  \rightarrow  f_{n}} \sum_{n'}\left( \frac{\mu^{[t-1]}_{f_{n'} \rightarrow h_n^{(k)}}}{\sigma^{2[t-1]}_{f_{n'} \rightarrow h_n^{(k)}}} \right).\\
    \end{split}
    \label{parameters_h_to_f_imperfec_chan_final}
\end{equation}

\subsection{Message from FN $f_n$ to VN $x^{(k)}$}
This is given by 
\begin{equation}
\begin{split}    
    & \lambda_{f_{n} \rightarrow x^{(k)}}^{[t]}(x^{(k)}) = 
   \int_{\mathbf{x}_{\sim x^{(k)}} \in \mathcal{C}^{K-1}, \mathbf{h}_n \in \mathbb{C}^K} 
    f_n\biggl( \mathbf{x},\mathbf{h}_n ; y_n\biggr)\\
   & \prod_{k'\neq k} \lambda_{x^{({k'})} \rightarrow f_{n}}^{[t-1]}(x^{(k')})  \prod_{k} \lambda_{h_n^{({k'})} \rightarrow f_{n}}^{[t-1]}(h_n^{(k')}) d\mathbf{x}_{\sim x^{(k)}} d\mathbf{h}_n, \\
\end{split}    \label{eq:message_from_function_node_x_to_variable_node_imperfect}
\end{equation}
where the integration is over $K-1$ information symbols and $K$ channel nodes. Using the same technique as in Sec. \ref{message_f_to_x_sec} to solve this integral, the message becomes $\lambda_{f_{n} \rightarrow x^{(k)}}^{[t]}(x^{(k)}) = \text{vM}\left(x^{(k)};\eta_{f_{n} \rightarrow x^{(k)}}^{[t]}\right)$, where 
\begin{equation}
\begin{split}
    & \eta_{f_{n} \rightarrow x^{(k)}}^{[t]} = \\ & \frac{2(\mu_{h_n^{({k})} \rightarrow f_{n}}^{[t-1]})^*}{\sigma^2}\left(y_n  - \sum_{{k'}\neq k}m_1(\eta_{ x^{({k'})}\rightarrow f_n}^{[t-1]}) \mu_{h_n^{({k'})} \rightarrow f_{n}}^{[t-1]}  \right).\\
\end{split}
\end{equation}

\subsection{Message from FN $f_n$ to VN $h_n^{(k)}$}
This is given by 
\begin{equation}
\begin{split}    
    & \lambda_{f_{n} \rightarrow h_n^{(k)}}^{[t]}(h_n^{(k)}) = 
   \int_{{\mathbf{h}_n}_{\sim h_n^{(k)}} \in \mathbb{C}^{K-1}, \mathbf{x} \in \mathcal{C}^K} 
   f_n\biggl( \mathbf{h}_n,\mathbf{x} ; y_n\biggr)\\
   & \prod_{k'\neq k} \lambda_{h_n^{({k'})} \rightarrow f_{n}}^{[t-1]}(h_n^{(k')})  \prod_{k} \lambda_{x^{({k'})} \rightarrow f_{n}}^{[t-1]}(x^{(k')}) d\mathbf{x} d{\mathbf{h}_n}_{\sim h_n^{(k)}}, \\
\end{split}    \label{eq:message_from_function_node_x_to_variable_node}
\end{equation}
where the integration is over $K-1$ channel nodes and $K$ information symbol nodes. Using the same technique as in Sec. \ref{message_from_x_to_f_sec} to solve this integral, the message becomes 
\begin{equation}
    \lambda_{f_{n} \rightarrow h_n^{(k)}}^{[t]}(h_n^{(k)}) = \text{CN} \left(h_n^{(k)};\mu_{f_{n} \rightarrow h_n^{(k)}}^{[t]}, \sigma_{f_{n} \rightarrow h_n^{(k)}}^{2[t]}\right),
\end{equation}
where the parameters are written as 
\begin{equation}
    \begin{split}
        &\mu_{f_{n} \rightarrow h_n^{(k)}}^{[t]} \\
        & = m_1(\eta_{ x^{({k})}\rightarrow f_n}^{[t-1]}) \left( y_n^* - \sum_{k'\neq k}m_1^*(\eta_{ x^{({k'})}\rightarrow f_n}^{[t-1]}) \mu_{ h_n^{({k'})}\rightarrow f_n}^{*[t-1]}\right),\\
        & \sigma_{f_{n} \rightarrow h_n^{(k)}}^{2[t]} = \sigma^2.
    \end{split}
\end{equation}

\subsection{Beliefs and Estimation}
The beliefs for VN $x^{(k)}$ are calculated using \eqref{eta_MIMO} and the corresponding estimation of symbols from \eqref{estimation_x}. Similarly, the belief of VN $h_n^{(k)}$ can be computed using 
\begin{equation}
    b^{[t]}(h_n^{(k)}) \propto f_{H_n^{(k)}}(h_n^{(k)})  \prod_{n'} \lambda_{f_{{n}} \rightarrow h_n^{(k)}}^{[t]}(h_n^{(k)}),
    \label{belief_chan}
\end{equation}
where $n'$ takes all values in $\{1,\cdots,N\}$ that are connected to the $h_n^{(k)}$ node.
Since all terms are CN distribution, the belief of VN $h_n^{(k)}$ is also a CN distribution, written as $b^{[t]}(h_n^{(k)}) = \text{CN} \left( h_n^{(k)}; \mu_{h_n^{(k)}}^{[t]},\sigma_{h_n^{(k)}}^{2[t]}\right)$, where the parameters are given by 
\begin{equation}
    \begin{split}
        & \sigma^{2[t]}_{h_n^{(k)}} = \left(\frac{1}{\sigma_h^{2[0]}} + \sum_{n'} \frac{1}{\sigma^{2[t]}_{f_{n'} \rightarrow h_n^{(k)}}}\right)^{-1},\\
        & \mu^{[t]}_{h_n^{(k)}} = \sigma^{2[t]}_{h_n^{(k)}} \left(\frac{\mu^{[0]}_{h_n^{(k)}}}{\sigma^{2[0]}_{h}} + \sum_{n'} \frac{\mu^{[t]}_{f_{n'} \rightarrow h_n^{(k)}}}{\sigma^{2[t]}_{f_{n'} \rightarrow h_n^{(k)}}} \right).\\
    \end{split}
\end{equation}
The channel state is estimated from the belief and the estimated symbol is given by 
\begin{equation}
    \hat{h}_n^{(k)} = \mu^{[N_{itr}]}_{h_n^{(k)}}.
\end{equation}

\section{Performance Analysis}
\label{sec_performance}
\subsection{Complexity for Coherent Case}

Let us first evaluate the complexity of the proposed algorithm in terms of floating point operations (FLOPs) per iteration, following the per-operation complexity from \cite[Table 1]{complexity}. 
In vM-BP, instead of tracking a mean and a variance as usual in approximate inference algorithms, message updates can be derived using only the complex natural parameter $\eta$.

The naive complexity number obtained by multiplying the respective complexities of computing \eqref{eq:natural_parameter_vn_to_fn} and \eqref{eq:natural_parameter_function_variable_final} by the number of their occurrences at each iteration of the algorithm  ($KN$ times, one message per edge)
can be reduced by noticing that the sums of all-but-one terms in \eqref{eq:natural_parameter_vn_to_fn} and \eqref{eq:natural_parameter_function_variable_final} which must be computed for all $k$ and $n$ have many terms in common. 
Computing the total sum once per iteration, either per observation or per user and  removing the contribution from the variable of interest to compute each message leads to a complexity of $33KN-10N-2K+KNC_A$ FLOPS, where
$C_A$ is the cost of computing $A(\cdot)$.
The efficient evaluation of $A(\cdot)$ is a rich research topic \cite{segura2023_bounds_ratio_modified_Bessel} which we do not address in this article; note however that it is a well-behaved scalar function of a scalar parameter, which can be approximated efficiently.
Hence we keep $C_{A}$ undefined for the purpose of complexity evaluation.
Note that the complexity of vM-BP is independent of the modulation order $M$ since the relaxation to the unit circle obliterates the need to consider the discrete modulation $\Omega$ during the iterations.

In Table \ref{tab:flops_comparison}, the complexity of the proposed vM-BP is compared to three other iterative algorithms: EP (for which we also assume the use of the aforementioned complexity reduction approach) and generalized-AMP (G-AMP) \cite{wu2014low}, which are both based on the same factor graph representation as vM-BP, as well as VEP \cite{senst2011framework}, in which all receive antennas are handled by a single factor node. 
VEP generally performs better at the cost of increased computational cost because each iteration involves a matrix inversion.

\begin{table}[!ht]
\centering
\caption{Total number of FLOPs per iteration}
\begin{tabular}{|c|c|}
\hline
\textbf{Algorithm} & \textbf{FLOPs per iteration} \\
\hline
vM-BP
& $33KN-10N-2K+KNC_A$ \\
\hline
EP 
& $37 KN + K M + 3 K$ \\
\hline
G-AMP 
& $32 K N + K M + 8 N + 5K$ \\
\hline 
VEP 
& $ \frac{2}{3} N^3+ 6 K N^2+ 5 K N+ K M+ 5K$ \\
\hline
\end{tabular}
\label{tab:flops_comparison}
\end{table}

\begin{figure}[h]
    \centering
    \resizebox{0.5\textwidth}{!}{
    \input{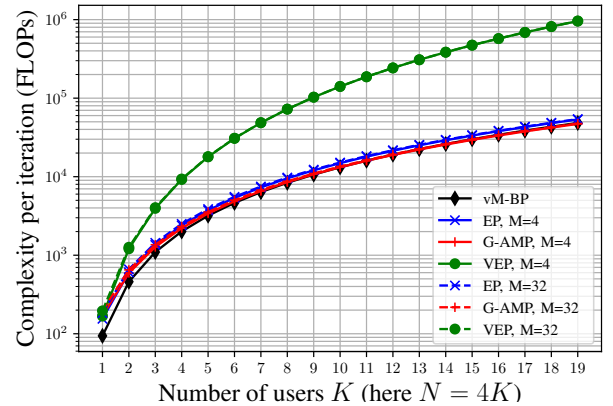}
    }
    \caption{Complexity comparison for different values of $K$ and $M$.}
    \label{fig:complexity}
\end{figure}

In Fig. \ref{fig:complexity}, the number of FLOPs is plotted against the number of users $K$ with the number of antennas set as $N = 4K$ (lightly loaded system), for different values of the PSK order $M$. Different iterative algorithms like VEP, EP, GAMP and vM-BP are compared. It can be seen that VEP is more complex compared to the rest of the decoders.  

\subsection{Error rate analysis}
In this section, we evaluate the performance of the proposed vM-BP for coherent MIMO detection in terms of symbol error rate (SER) for different system parameters. We normalize each channel vector in \eqref{MIMO_rx_signal} such that $||\mathbf{h}^{(k)}||=1$ for all $k$, and first assume exact knowledge of the channel state. The signal to noise ratio (SNR) is defined as $1/\sigma^2$.
For simulations, a system with  $K=16$ users  and $N=64$ receive antennas is considered.
The proposed vM-BP algorithm is compared with L-MMSE, EP, AMP and VEP.
In this scenario with $K\ll N$, we consistently observed that damping had almost no impact, hence it is not considered in this section.

\begin{figure}[h]
    \centering
    \resizebox{0.5\textwidth}{!}{
    \input{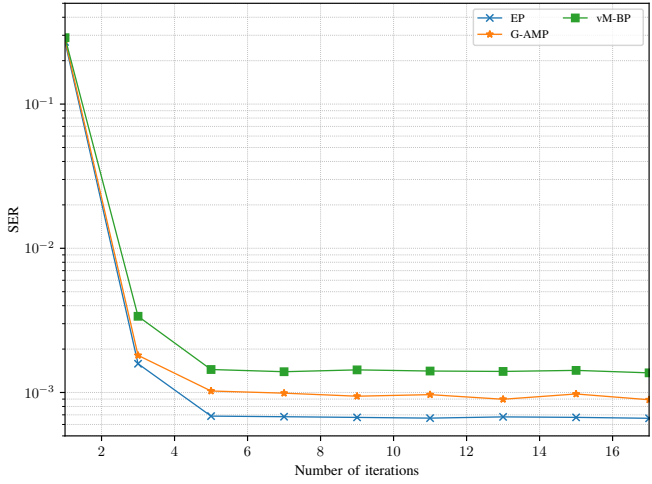}
    }
    \caption{SER versus number of iterations for Rayleigh fading channel with $K = 16, N = 64, M = 8$, SNR=16 dB.}
    \label{fig:ser_iteration}
\end{figure}

Let us first focus on the convergence behavior. Fig.~\ref{fig:ser_iteration} depicts the SER versus number of iterations for SNR = 16 dB and $M=8$ for vM-BP, G-AMP and EP, for Rayleigh fading. 
All algorithms require about $5$ iterations to converge. 
EP has the lowest plateau with a minimal SER equal to $7.10^{-4}$, G-AMP stops at $9.10^{-4}$ and vM-BP at $10^{-3}$. 
In the sequel, the number of iterations is set to $10$ to ensure convergence.

Performance in terms of SER versus SNR is evaluated over two different channel models: a) Gaussian i.i.d. channel and b) Correlated channel modeled using angle of arrivals (AoA).

\subsubsection{Gaussian i.i.d. Channels}
\begin{figure}[h]
    \centering
    \resizebox{0.5\textwidth}{!}{%
        \input{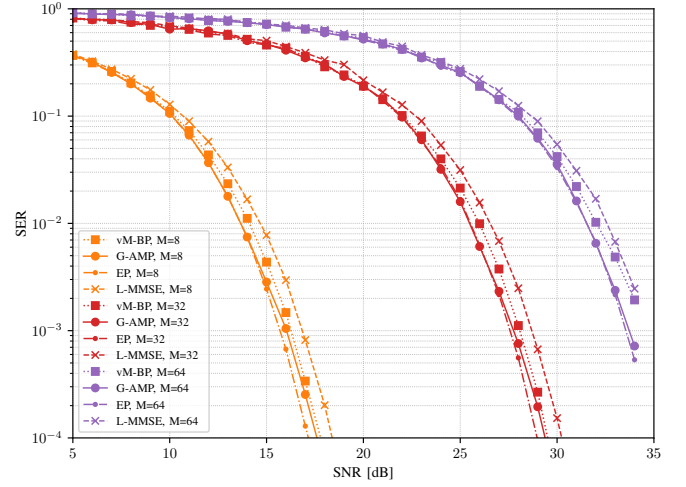}
    }
    \caption{SER for Rayleigh fading channel with $K = 16, N = 64, M = 8, 32, 64$.}
    \label{fig:ser_snr_iid_2}
\end{figure}

Fig. \ref{fig:ser_snr_iid_2} shows the SER against SNR for different $M$ assuming that the receiver has perfect knowledge of the true Gaussian i.i.d. (Rayleigh fading) channel.
The performance of VEP is not depicted because it performs identically to G-AMP in this scenario.
The three iterative algorithms outperform L-MMSE detection for all considered modulation orders ($M=8$, 32, 64). While AMP outperforms vM-BP at low SNR, which is coherent with Fig. \ref{fig:ser_iteration}, the gap becomes smaller as the SNR increases. Scalar EP consistently outperforms vM-BP by $0.5$ dB.

\subsubsection{Correlated channels}
In this section, the channels follow a geometric model assuming line-of-sight propagation and a uniform linear array with half-wavelength spacing at the receiver side, hence the channel of each user consists in the array steering vector which is fully determined by the angle of arrival (AoA). We consider three different scenarios, which differ in how the AoAs are chosen: (i) high correlation, with AoAs for user $k$ given as $k\degree$ from the broadside direction, where the channel matrix $\mathbf{H}$ has condition number as $4.6 \cdot 10^4$, (ii) medium correlation with AoAs linearly spaced between $[-10\degree, 10\degree]$, with channel condition number $201$, and (iii) low correlation, with AoAs linearly spaced in $[-60\degree, 60\degree]$ for which the channel condition number is $1.2$.

\begin{figure}[t]
    \centering
    \resizebox{0.5\textwidth}{!}{%
        \input{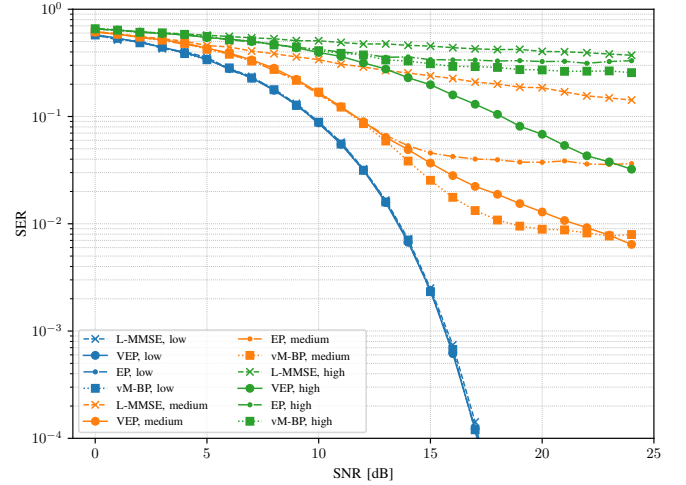}
    }
    \caption{SER for high, medium and low correlation channels, $K = 16, N = 64, M = 8$.}
    \label{fig:ser_snr_corr_2}
\end{figure}

The performance curves depicted in Fig.~\ref{fig:ser_snr_corr_2} show that, for all three cases, vM-BP performs better than L-MMSE.
As expected, increased correlation (and channel condition number) is detrimental to all detection algorithms.
All detectors exhibit the same performance for the low correlation scenario. This is expected since in LoS conditions with hardly any correlation with the other users the detection is easy.
The performance of G-AMP is not depicted, it is essentially identical to that of EP in this scenario.
In the high correlation scenario, only VEP is able to detect somewhat reliably.
The results for the medium correlation are different. The error floor for vM-BP, at $8.10^{-3}$, is lower than the one for scalar EP, at $3.10^{-2}$. Also, although  VEP does not suffer from an error floor, it is outperformed by vM-BP from 13 dB to 23 dB. 
No intuitive explanation is available for the discrepancy between the performance rankings of the various algorithms over high and medium correlated channels.

\subsubsection{Overloaded System with Gaussian i.i.d. Channel}
\begin{figure}[t]
    \centering
    \resizebox{0.5\textwidth}{!}{%
        \input{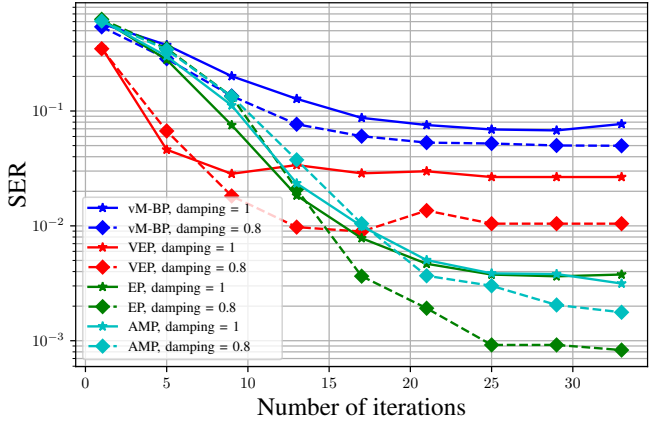}
    }
    \caption{\centering SER versus number of iterations for overloaded system with $K = 120, N = 96, M = 8$.}
    \label{ser_itr_overloaded}
\end{figure}

\begin{figure}[t]
    \centering
    \resizebox{0.5\textwidth}{!}{%
        \input{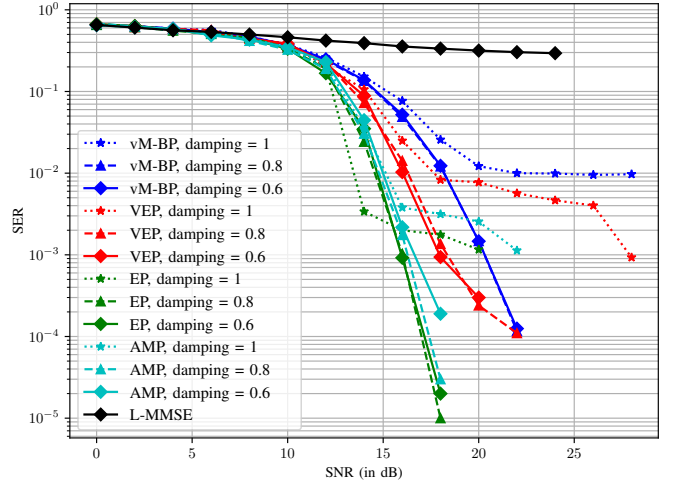}
    }
    \caption{\centering SER for overloaded system with Gaussian i.i.d. fading, $K = 120, N = 96, M = 8$.}
    \label{ser_overloaded}
\end{figure}

We consider an overloaded system with $K = 120$ users and $N = 96$ receive antennas, for PSK of order $M = 8$, with Gaussian i.i.d. channel. The SER versus number of iterations for SNR = 16 dB is plotted in Fig. \ref{ser_itr_overloaded} for different damping factors. This damping is introduced in the messages from FN to VN and an improvement in performance is noted. All algorithms seem to converge at around 25 iterations. Hence we fix the number of iterations to 25 for all algorithms in the next simulation, depicted in Fig. \ref{ser_overloaded} where SER is plotted against SNR.
For Fig. \ref{ser_overloaded}, the performance is compared with L-MMSE, G-AMP, VEP and EP. As expected, the iterative algorithms significantly outperforms L-MMSE, which as a linear method, is not able to resolve the over-determined system of equations even at high SNR. From Fig. \ref{ser_overloaded} it is evident that the iterative algorithms do not perform well with no damping and an improvement in performance is noticed when damping is introduced.

\subsection{Imperfect Channel Knowledge}
\begin{figure}[t]
    \centering
    \resizebox{0.5\textwidth}{!}{%
        \input{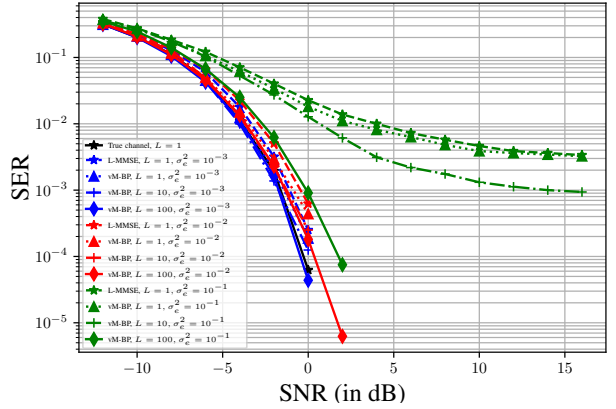}
    }
    \caption{\centering SER for imperfect channel with $K = 16, N = 64, M = 8$.}
    \label{ser_imperfect_chan}
\end{figure}

In this subsection, we analyze the performance of the algorithm from Sec.~\ref{sec_imperfectCSI} when only imperfect CSI is available.
The imperfect channel is modeled as  
\begin{equation}
    \tilde{h}_n^{(k)} = h_n^{(k)} + e,
    \label{imperfect_chan_mod}
\end{equation}
where the true channel $h_n^{(k)}$ is Gaussian i.i.d. (with no normalization) and $e$ is a complex channel estimation error with zero mean and variance $\sigma_e^2$. 
We consider three scenarios: (i) a mismatched detector, where the receiver assumes that the true CSI is equal to $\tilde{h}_n^{(k)}$ and works on the factor graph from Fig.~\ref{factor_graph_MIMO} to estimate only the information symbols; (ii) a CSI-imperfection-aware detector where the receiver knows that $\tilde{h}_n^{(k)}$ is imperfect CSI and is aware of the estimation error variance $\sigma_e^2$, and tries to obtain an improved estimate of CSI along with information symbols, using the factor graph from Fig.~\ref{factor_graph_MIMO_imperfect_chan}; (iii) a block-fading scenario with (known) blocklength $L$, whereby each user transmits $L$ symbols while the channel remains unchanged; the receiver tries to estimate both the symbols and channel. This is an extension of scenario (ii) where the graph is extended by adding information symbol nodes.
SER performance curves are depicted in Fig.~\ref{ser_imperfect_chan}. 
For the mismatched decoder case we plot L-MMSE performance, while for the imperfection-aware decoder we plot vM-BP. For $L = 1$, i.e., each user transmitting only one symbol, damping did not improve the performance. However, for $L > 1$, damping showed significant improvement in performance, and we fix the damping coefficient to 0.6 for messages from FN to VN (both channel and information symbol nodes). The number of iterations for this scenario is set to 20. 
It can be noted that, the imperfection-aware detector improves the performance. As the number of symbols transmitted from each user increase, SER decreases, as the channel estimates are improved since the degree of channel node increases. As $L$ increases, the SER performance approaches the performance achievable under perfect CSI, even when provided with a highly imperfect prior (high $\sigma_e^2$).

\section{Conclusion}
\label{sec_conclusion}
This paper proposed a BP detector for mMIMO with PSK modulations based on von Mises distributions by leveraging directional statistics and relaxing the discrete constellation on the unit circle to a continuous vM distribution.
This results in an algorithm that is fully continuous and agnostic to the modulation order. 
The numerical results and complexity of the algorithm are discussed in detail and compared to state of the art iterative detectors, such as EP, G-AMP. Extensions of the proposed algorithm to the presence of imperfect channel are also presented, making the proposed algorithm suitable for practical mMIMO systems.

\bibliographystyle{IEEEtran}
\bibliography{bibliography} 

\end{document}